\documentclass[aps,pra,twocolumn]{revtex4-2}
\usepackage[T1]{fontenc}
\usepackage{bm}
\usepackage{amsmath}
\usepackage{amssymb}
\usepackage{slashed}
\usepackage{float}
\allowdisplaybreaks
\usepackage{dcolumn}
\newcolumntype{d}[1]{D{.}{.}{#1}}
\newcommand*{\centt}[1]{\multicolumn{1}{c}{#1}}

\begin{document}

\title{Nuclear mass and size corrections to the magnetic shielding}
\author{Krzysztof Pachucki}
\affiliation{Faculty of Physics, University of Warsaw,
             Pasteura 5, 02-093 Warsaw, Poland}

\begin{abstract}
We derive finite nuclear mass and finite nuclear size corrections to the magnetic shielding in light ions.
These corrections are important for the accurate determination of nuclear magnetic moments.
We correct several previous  formulas for the nuclear mass corrections and present improved results  for the magnetic shielding in 
$^1$H, $^3$He$^+$, and $^3$He. 
Finally, we obtain an $^3$He atomic magnetic moment, which serves as an accurate probe to measure magnetic fields.
\end{abstract}

\maketitle

\section{Introduction}
The nuclear magnetic moment in atoms is partially shielded by atomic electrons.
This effect is not very significant: about $10^{-5}$ for light elements. Nevertheless, because
nuclear magnetic moments are determined from the Zeeman shift in atomic systems,
the calculation of the magnetic  shielding is necessary for their accurate determination.
For example, the recent measurement of the magnetic moment of the $^3$He$^+$ ion \cite{schneider2022} 
together with the calculation of magnetic shielding \cite{wehrli:2021, wehrli:2022} 
allowed  for the most accurate determination so far of the helion magnetic moment.
A similar measurement is planned for $^9$Be$^{3+}$, which will result in an improved determination
of the $^9$Be nuclear magnetic moment \cite{private3}.
Moreover, accurate values for nuclear magnetic moments are important for the determination of atomic hyperfine splitting (HFS), 
testing quantum electrodynamics (QED), and the nuclear structure theory \cite{SGK05}. 
This is because HFS is very sensitive to the distribution of the magnetic moment within the nucleus.

In this work, we point out two interesting effects that are frequently overlooked in calculations
of nuclear magnetic shielding \cite{jaszunski2012}, namely, those due to the finite nuclear mass and the finite nuclear size. 
Nuclear mass corrections are as large as relativistic corrections for light atomic systems,
while finite nuclear size effects are much smaller, but they are expected to be significant for heavier elements.
These finite nuclear mass effects have already been the subject of several works \cite{pachucki08, rudzinski09}. 
Here, we rederive them thoroughly, correct some mistakes, and update numerical values for the most relevant cases of 
the H, $^3$He$^+$, and $^3$He elements. 
The finite nuclear size effects have been studied only numerically and only for hydrogen-like systems \cite{yerokhin11,yerokhin12}.
Here, we derive a compact analytic formula in terms of the charge, magnetic, and effective Zemach nuclear radii,
which accounts also for nuclear inelastic effects.

\section{Breit-Pauli Hamiltonian with the homogenous magnetic field}
To account for finite nuclear mass effects, we have to treat nuclei on an equal footing with all electrons.
Therefore, we consider a system of charged particles, each having its own mass $m_a$, charge $e_a$, spin $s_a$, 
and the so-called g-factor $g_a$, which is related to the magnetic moment by
\begin{align}
\vec \mu_a = \frac{g_a\,e_a}{2\,m_a}\,\vec s_a\,. \label{02}
\end{align} 
These particles are electrons with spin $1/2$ and nuclei with an arbitrary spin.
Our derivation employs a Breit-Pauli Hamiltonian with homogenous magnetic field 
and with separation of center of mass motion. It closely follows the lines of Ref. \cite{pachucki08}. 
Let us therefore  introduce the total mass $M$
\begin{align}
M = \sum_a m_a\,, \label{03}
\end{align}
center of mass variables
\begin{eqnarray}  
\vec R &=& \sum_a \frac{m_a}{M}\,\vec r_a\,,\label{04}\\
\vec P &=& \sum_a \vec p_a\,,\label{05}
\end{eqnarray}
and relative coordinates
\begin{eqnarray}
\vec x_a &=& \vec r_a-\vec R\,, \label{06}\\
\vec q_a &=& \vec p_a-\frac{m_a}{M}\,\vec P\,, \label{07}
\end{eqnarray}
such that
\begin{eqnarray}
\bigl[x_a^i\,,\,q_b^j\bigr] &=&
i\,\delta^{ij}\,\biggl(\delta_{ab}-\frac{m_b}{M}\biggr)\,, \label{08}\\
\bigl[R^i\,,\,P^j\bigr] &=& i\,\delta^{ij}\,,  \label{09}\\
\bigl[x_a^i\,,\,P^j\bigr] &=& \bigl[R^i\,,\,q_a^j\bigr] = 0\,. \label{10}
\end{eqnarray}
The Hamiltonian of a bound system of charged particles in an external magnetic field including leading relativistic corrections
and with the separated-out center of mass motion  is \cite{pachucki08}
\begin{widetext}
\begin{align}
H_\mathrm{in} =&\ \sum_a\biggl\{\frac{\vec \pi_a^2}{2\,m_a} -\frac{e_a}{2\,m_a}\,g_a\,\vec s_a\cdot\vec B
-\frac{\vec\pi^4_a}{8\,m^3_a}
+\frac{e_a}{8\,m^3_a}\,\Bigl[
4\,\vec\pi_a^2 \, \vec s_a\cdot\vec B
+(g_a-2)\,\bigl\{\vec\pi_a\cdot\vec B\,,\,\vec\pi_a\cdot\vec s_a\bigr\}\Bigr] -\frac{e_a^2}{2}\,\chi_{a}\,\vec B^2\biggr\}
\nonumber \\&\ 
+\sum_{a>b,b} \frac{e_a\,e_b}{4\,\pi}\,\biggl\{ \frac{1}{r_{ab}}
-\frac{1}{2\,m_a\,m_b}\,\pi_a^i\,
\biggl(\frac{\delta^{ij}}{r_{ab}}+\frac{r^i_{ab}\,r^j_{ab}}{r^3_{ab}}
\biggr)\,\pi_b^j -
\frac{2\,\pi}{3}\,\langle r_{Ea}^2+r_{Eb}^2\rangle \, \delta^3(r_{ab})
-\frac{2\,\pi\,g_a\,g_b}{3\,m_a\,m_b}\,\vec s_a \cdot\vec s_b\,\delta^3(r_{ab})
\nonumber \\ &\  
+\frac{g_a\,g_b}{4\,m_a\,m_b}\,\frac{s_a^i\,s_b^j}
{r_{ab}^3}\,
\biggl(\delta^{ij}-3\,\frac{r_{ab}^i\,r_{ab}^j}{r_{ab}^2}\biggr)\biggr\}
+\sum_{a,b} \frac{e_a\,e_b}{4\,\pi}\,\frac{1}{2\,r_{ab}^3} \biggl[
\frac{g_a}{m_a\,m_b}\,\vec s_a\cdot\vec r_{ab}\times\vec\pi_b 
 -\frac{(g_a-1)}{m_a^2}\,\vec s_a\cdot\vec
r_{ab}\times\vec\pi_a\biggr]\,, \label{11}
\end{align}  
\end{widetext}
where $r_{ab} = |\vec r_a-\vec r_b|$,  and
\begin{align}
\vec \pi_a =&\ \vec q_a +\frac{1}{2}\,\vec D_a\times\vec B\,, \label{12}\\
\vec D_a =&\ e_a\,\vec x_a+\frac{m_a}{M}\,\vec D\,, \label{13}\\
\vec D =&\ \sum_a e_a\,\vec x_a\,. \label{14}
\end{align}
For a point spin $s=1/2$ particle, $g=2$, $\langle r_E^2\rangle =3/(4\,m^2)$, and $\chi = 1/(4\,m^3)$.
For a finite size particle $\langle r_E^2\rangle$ includes the mean square charge radius.
An equivalent Hamiltonian for a system of spin $1/2$ point particles was originally obtained by Hegstrom in Ref. \cite{hegstrom73}.
Our Hamiltonian in Eq. (\ref{11}), however,  is valid for arbitrary spin particles and has a more compact form. 

The magnetic interaction resulting from $H_\mathrm{in}$ neglecting the terms quadratic in $\vec B$ is
\begin{align}
\delta H =& -\sum_a\frac{e_a}{2\,m_a}\,(\vec x_a\times\vec q_a + g_a\,\vec s_a)\cdot\vec B
\nonumber \\ &
+\sum_a\,\frac{1}{4\,m_a^3}\,\Bigl[
q_a^2\,\vec D_a\times\vec q_a\cdot\vec B
+2\,e_a\,q_a^2\,\vec s_a\cdot\vec B
\nonumber \\ &
+e_a\,(g_a-2)\,\vec q_a\cdot\vec s_a\,\vec q_a\cdot\vec B\Bigr]
+\sum_{a \neq b,b}\,\frac{e_a\,e_b}{4\,\pi}\biggl[
\nonumber \\ &
-\frac{1}{4\,m_a\,m_b}\,q_a^i\biggl(\frac{\delta^{ij}}{r_{ab}}+
\frac{r_{ab}^i\,r_{ab}^j}{r_{ab}^3}\biggr)\,(\vec D_b\times\vec B)^{\,j}
\nonumber \\ &
+\frac{1}{4\,r_{ab}^3}\,\frac{g_a}{m_a\,m_b}\,
(\vec s_a\times\vec r_{ab})\cdot(\vec D_b\times\vec B)
\nonumber \\ &
-\frac{1}{4\,r_{ab}^3}\,\frac{(g_a-1)}{m_a^2}\,
(\vec s_a\times\vec r_{ab})\cdot(\vec D_a\times\vec B)\biggr].
\label{15}
\end{align}
This is a general interaction Hamiltonian, which is valid for an arbitrary set of particles.
In particular, one can obtain the bound electron g-factor or the magnetic shielding in
atomic and molecular systems.  In the next section we derive the atomic magnetic shielding
with full account of the nuclear mass.

\section{Finite nuclear mass corrections}
We will derive the magnetic shielding constant for arbitrary ions with the vanishing orbital angular momentum $\vec L$. 
The interaction of the nuclear spin with the magnetic field is
obtained from Eq. (\ref{15}) as
\begin{align}
\delta H =&\ -\frac{e_N}{2\,m_N}\,g_N\,\vec s_N\cdot\vec B
\nonumber \\ &
+\frac{e_N}{4\,m_N^3}\,\Bigl[
2\,q_N^2\,\vec s_N\cdot\vec B
+(g_N-2)\,\vec q_N\cdot\vec s_N\;\vec q_N\cdot\vec B\Bigr]
\nonumber \\ &\ 
+\sum_{b}{}'\,\frac{e_N\,e}{4\,\pi}\,
\frac{(\vec s_N\times\vec r_{Nb})}{4\,r_{Nb}^3}
\nonumber\\ &\ 
\cdot\biggl[\frac{g_N}{m_N\,m_e}\,
\vec D_b\times\vec B-\frac{(g_N-1)}{m_N^2}\,\vec D_N\times\vec B\biggr],
\label{16}
\end{align}
where we assumed that for nucleus $a=N$, all other particles are electrons, and $\sum'$ denotes summation over electrons only.
For an electronic state with a spherical symmetry
\begin{align}
(\vec s_N\times\vec X)\cdot(\vec Y\times\vec B) =-\frac{2}{3}\,\vec s_N\cdot\vec B\,\vec X\cdot\vec Y\,, \label{17}
\end{align}
one introduces the scalar shielding constant $\sigma$
\begin{align}
\delta H =  -\frac{g_N\,e_N}{2\,m_N}\,\vec s_N\cdot\vec B(1-\sigma)\,. \label{18}
\end{align}
This $\sigma$ is conveniently split into two parts, consisting of the first- and the second-order matrix elements
\begin{align}
\sigma = \sigma_1+\sigma_2\,. \label{19}
\end{align}
$\sigma_1$ results from the first-order matrix element of $\delta H$ in Eq. (\ref{16})
\begin{align}
\sigma_1 =&\ \frac{1}{2\,g_N\,m_N^2}\,\Bigl[2+\frac{(g_N-2)}{3}\Bigr]\,\big\langle q_N^2\big\rangle \label{20}\\ &
+\frac{1}{3}\,\frac{e_e}{4\,\pi}\bigg\langle \sum_{b}{}'\,\frac{\vec r_{bN}}{r_{bN}^3}
\cdot\biggl[\frac{1}{m_e}\,\vec D_b-\frac{(g_N-1)}{g_N\,m_N}\,\vec D_N\biggr]\bigg\rangle. \nonumber
\end{align}
Because $\sum_a m_a\,\vec x_a=0$, the position  $\vec x_N$ of the nucleus with respect to mass center and the dipole operator $\vec D$
can be expressed in terms of the electron coordinates only
\begin{align}
\vec x_N =&\ -\frac{m_e}{M}\,\sum_a{}'\,\vec r_{aN}\,, \label{21}\\
\vec D =&\ e_e\,\sum_a{}' \vec r_{aN}\,\Bigl(1+(Z-N_e)\,\frac{m_e}{M} \Bigr)\,, \label{22}
\end{align}
where $M=m_N + N_e\,m_e$, $N_e$ is the number of electrons, and $Z$ is the nuclear charge in units of the elementary charge. 
Consequently,  the shielding constant $\sigma_1$ takes the form
\begin{align}
\sigma_1 =&\  \frac{\alpha}{3\,m_e}\,\biggl\langle\sum_{a}{}'\,\frac{1}{r_{a}}\biggr\rangle + \frac{(4+g_N)}{6\,g_N}\,\frac{\langle p_N^2\rangle}{m_N^2}
\nonumber \\ &\
+\frac{\alpha}{3}\,\bigg\langle\sum_{b}{}'\,\frac{\vec r_{bN}}{r_{bN}^3}\cdot\sum_a{}' \vec r_{aN}\bigg\rangle\,
\frac{1}{g_N}\,\frac{m_e}{M}\,
\nonumber \\ &\times
\biggl[\frac{(Z-N_e)}{M}  +(1-g_N)\,\bigg(\frac{1}{m_e} + \frac{Z}{m_N}\bigg)\biggr]\,,
\label{23}
\end{align}
where $\vec p_N = -\sum_a{}' \vec p_a$, and we used $\vec P|\phi\rangle = 0$.
Consider now the following matrix element
\begin{align}
\biggl\langle\sum_{a}{}'\,\frac{\vec r_{aN}}{r_{aN}^3}\cdot \sum_b{}' \vec r_{bN}\biggr\rangle 
=&\ \frac{1}{i\,Z\,\alpha}\,\biggl\langle[\vec p_N\,,\,H-E] \sum_b{}' \vec r_{bN}\biggr\rangle \nonumber \\ 
=&\ \sum_b{}'\frac{1}{i\,Z\,\alpha}\,\bigl\langle\vec p_N\,[H-E\,,\, \vec r_{bN}]\bigr\rangle \nonumber \\ 
=&\ \langle p_N^2\rangle\,\frac{1}{Z\,\alpha}\,\frac{M}{m_N\,m_e}\,,
\label{24}
\end{align}
which is used to simplify $\sigma_1$
\begin{align}
\sigma_1 =&\ 
\frac{\alpha}{3\,m_e}\,\bigg\langle\sum_{a}{}'\,\frac{1}{r_{a}} \bigg\rangle
+\frac{\langle p_N^2\,\rangle}{3\,g_N\,m_N^2}\,\biggl[3-\frac{g_N}{2} 
\nonumber \\ &\
+\frac{m_N}{M}\,\biggl(1-\frac{N_e}{Z}\biggr)  +(1-g_N)\,\frac{m_N}{Z\,m_e} \biggr]\,.
\label{25}
\end{align}

The $\sigma_2$ part is given by the second-order interaction coming from the Hamiltonian
\begin{align}
\delta H =&\ -\sum_a\frac{e_a}{2\,m_a}\,\vec x_a\times\vec q_a\cdot\vec B\,
-\frac{e_N\,e_e}{4\,\pi}\,\frac{\vec s_N}{2\,m_N}
\nonumber \\ &
\cdot \sum_{b}{}' \frac{\vec r_{bN}}{r_{bN}^3}\times
\biggl[g_N\,\frac{\vec q_b}{m_e}-(g_N-1)\,\frac{\vec q_N}{m_N}\biggr],
\label{26}
\end{align}
namely
\begin{align}
\sigma_2 =&\frac{2}{3}\,
\biggl\langle \sum_a\frac{e_a}{2\,m_a}\,\vec x_a\times\vec q_a
\,\frac{1}{(E-H)}\,
\sum_{b}{}'\,\frac{e_e}{4\,\pi}\,
\frac{\vec r_{bN}}{r_{bN}^3}
\nonumber \\ &\times
\biggl[\frac{\vec p_b}{m_b}-\frac{(g_N-1)}{g_N}\,\frac{\vec p_N}{m_N}\biggr]\biggr\rangle\,.
\label{27}
\end{align}
Using
\begin{align}
\sum_a \frac{e_a}{2\,m_a}\,\vec x_a\times \vec q_a 
=&\  \frac{e_e}{2\,m_e}\,\vec L + \biggl( \frac{e_e}{2\,m_e} + \frac{Z\,e_e}{2\,m_N}\biggr)\nonumber \\ &\times
\frac{m_e}{M}\,\sum_a{}' \vec r_{aN}\times\vec p_N,
\label{28}
\end{align}
we arrive at
\begin{align}
\sigma_2 =&
\frac{\alpha}{3\,M}\,\biggl(1 + \frac{Z\,m_e}{m_N}\biggr)\,
\biggl\langle 
\sum_a{}' \vec r_{aN}\times\vec p_N
\,\frac{1}{(E-H)}\,
\sum_{b}{}'\frac{\vec r_{bN}}{r_{bN}^3}
\nonumber \\ &\times
\biggl[\frac{\vec p_b}{m_e}-\frac{(g_N-1)}{g_N}\,\frac{\vec p_N}{m_N}\biggr]
\biggr\rangle\,. \label{29}
\end{align}

The total shielding constant is $\sigma=\sigma_1+\sigma_2$, where $\sigma_1$ is given in Eq. (\ref{25}) and $\sigma_2$ in Eq. (\ref{29}).
For the numerical calculations,  it is convenient to apply the expansion in the mass ratio, which takes the form (in a.u.)
 \begin{align}
\sigma =&\ \sigma^{(2,0)} + \sigma^{(2,1)} +\ldots , \label{30}\\ 
\sigma^{(2,0)} =&\  \frac{\alpha^2}{3}\,\biggl\langle\sum_a{}'\frac{1}{r_{a}}\biggr\rangle\,, \label{31}\\
\sigma^{(2,1)} =&\ \frac{\alpha^2}{3}\,\frac{m_e}{m_N}\biggl[
\biggl\langle\sum_a{}'\frac{1}{r_{a}}\frac{1}{(E-H)'}\,p_N^2\biggr\rangle
+\langle p_N^2\,\rangle \frac{(1-g_N)}{Z\,g_N}
\nonumber \\ &\
+\biggl\langle\sum_a{}'\vec r_a\times\vec p_N\,\frac{1}{(E-H)'}\,
\sum_{b}{}'\,\frac{\vec r_b\times\vec p_b}{r_b^3}\biggr\rangle\biggr],
\label{32}
\end{align}
where all matrix elements in the above are assumed with infinite nuclear mass, 
and $\sigma^{(i,j)}$ denotes the expansion term of order $\alpha^i\,(m_e/m_N)^j$.
The last term in the above differs in sign from that derived previously in Ref. \cite{pachucki08, rudzinski09},
see Table I for the updated numerical values.


For the hydrogenic ion in $nS$ state the nonrelativistic shielding constant, using Eq. (\ref{25}),  is
\begin{align}
\sigma =&\ \frac{(Z\,\alpha)^2}{3\,n^2}\,\frac{m_N}{m_N+m_e}+
\frac{(Z\,\alpha)^2\,m_e^2}{3\,n^2\,g_N\,(m_N+m_e)^2}
\nonumber \\ \times&
\biggl[ 3-\frac{g_N}{2} +\frac{m_N}{m_N+m_e}\,
\biggl(1-\frac{1}{Z}\biggr)  -(g_N-1)\,\frac{m_N}{Z\,m_e} \biggr] \nonumber \\
=&\ 
\frac{(Z\,\alpha)^2}{3\,n^2\,g_N\,(1+x)^2}\,\biggl[
\biggl(3-\frac{g_N}{2}+\frac{x}{1+x}\biggr)
\nonumber \\ &\
+\frac{x^2}{Z}\,\biggl(\frac{1}{1+x} +g_N\biggr)\biggr]\,,
\label{34}
 \end{align}
 where $x = m_N/m_e$. It is convenient to define  $\delta g_N = -g_N\,\sigma$
 \begin{align}
 \delta g_N =&\  \frac{(Z\,\alpha)^2}{3\,n^2\,(1+x)^2}\,\biggl[
\frac{g_N}{2} -3
-\frac{x}{1+x}
 \nonumber \\ &\ 
-\frac{x^2}{Z}\,\biggl(\frac{1}{1+x} +g_N\biggr)\biggr] \label{35}\,,
 \end{align}
which  is in agreement with the known formula for the electron g-factor in the S-state of the hydrogenic ion \cite{grotch71, pachucki08},
\begin{align}
\delta g_e =&\ \frac{(Z\,\alpha)^2}{3\,n^2\,(1+x)^2}\,\biggl[
\biggl(\frac{g_e}{2}-3-\frac{x}{1+x}\biggr)
 \nonumber \\ &\ 
-Z\,x^2\,\biggl(\frac{1}{1+x} + g_e\biggr)\biggr]
\label{36}
\end{align}
with $x=m_e/m_N$, which verifies the new formula for the magnetic shielding in hydrogen-like ions. 
Its small electron mass expansion takes the following form:
\begin{align}
\sigma =&\ 
\frac{Z\,\alpha^2}{3\,n^2}\,\biggl[
1 + \frac{m_e}{m_N}\,\bigg(\frac{1}{g_N}-2\bigg) 
 \nonumber \\ &\ 
+ \frac{m_e^2}{m_N^2}\,\bigg(\frac{4\,Z-3}{g_N} - \frac{Z}{2} + 3 \bigg) +\ldots
\biggr]\,, \label{37}
 \end{align}
where the quadratic in the mass ratio  term differs from that derived previously in Ref. \cite[Eq. (64)]{wehrli:2022}
due to the computational mistake.
As seen from Table I, the largest uncertainty for light ions comes from the relativistic
recoil correction $\sigma^{(4,1)}$, but this has not yet been studied in the literature.
 
 \section{Finite nuclear size corrections}
 Let us pass now to another nuclear correction, which is due to the finite distribution of
the charge and the magnetic moment within the nucleus.
We will study this correction for hydrogenic ions only, but generalization for an arbitrary ion is straightforward.
For light hydrogenic ions this effect is given by
  \begin{align}
  \sigma_\mathrm{fs} = -\frac{Z\,\alpha^2}{3}\,\bigl[
  2\,(Z\,\alpha)^2\,m^2\,(r_C^2+r_M^2) +8\,(Z\,\alpha)^3\,m\,\tilde r_Z \bigr], \label{38}
  \end{align}
  where $m=m_e$, $r_C$ is the charge radius, $r_M$ the magnetic radius, and $\tilde r_Z$ the effective Zemach radius of the nucleus.
 This formula is proved as follows.
  
  The shift of nonrelativistic hydrogenic levels due to $r_C$  is given by
  \begin{align}
  \delta H =&\ e\,A^0-\frac{e}{6}\,r_C^2\,\vec\nabla\vec E \nonumber \\
  =&\ -Z\,\alpha\,\Bigl( \frac{1}{r}-\frac{2\,\pi}{3}\,\delta^3(r)\,r_C^2\Bigr)\,, \label{39}
  \end{align}
  where $e=e_e$.
  The finite nuclear size affects the nonrelativistic wave function, which in turn affects the 
  matrix elements for the nuclear magnetic shielding
  \begin{align}
  \sigma_C =&\ 2\,\frac{\alpha}{3\,m}\,\Bigl\langle
 \frac{1}{r}\,\frac{1}{(E-H)'}\, \frac{2\,\pi}{3}\,Z\,\alpha\,r_C^2\, \delta^3(r)\Bigr\rangle
\nonumber \\
=&\ -2\,(Z\,\alpha)^2\,m^2\,r_C^2\,\frac{Z\,\alpha^2}{3}\,. \label{40}
\end{align} 

To derive the contribution from the magnetic radius of the nucleus, let us rederive the leading shielding
that comes from the $e^2\,\vec A^2/(2\,m)$ term in the kinetic energy of the electron 
\begin{align}
\delta E =&\ \frac{\alpha}{2\,m}\,\Bigl\langle(\vec B\times\vec r)\cdot\Bigl(\vec\mu\times\frac{\vec r}{r^3}\Bigr)\Bigr\rangle \nonumber \\
=&\ \vec\mu\cdot\vec B\,\frac{\alpha}{3\,m}\Bigl\langle\vec r\cdot\frac{\vec r}{r^3}\Bigr\rangle. \label{41}
\end{align}
The shielding $\sigma$ is thus given by
\begin{align}
\sigma =&\ -\frac{\alpha}{3\,m}\,\Bigl\langle \vec r\cdot\vec\nabla\Big(\frac{1}{r}\Big)\Bigr\rangle. \label{42}
\end{align}
The magnetic radius $r_M$ enters the magnetic interaction similarly to $r_C$ in Eq. (\ref{39});
therefore,  the shift due to the magnetic radius is
\begin{align}
\sigma_M=&\ 
\frac{\alpha}{3\,m}\,\frac{2\,\pi}{3}\,r_M^2\,\Bigl\langle \vec r\cdot\vec\nabla\Bigl(\delta^3(r)\Bigr)\Bigr\rangle \nonumber \\
=&\ 
\frac{\alpha}{3\,m}\,\frac{2\,\pi}{3}\,r_M^2\,(-3)\,\langle \delta^3(r)\rangle \nonumber \\
=&\ 
-2\,(Z\,\alpha)^2\,m^2\,r_M^2\,\frac{Z\,\alpha^2}{3}. \label{43}
\end{align}

The calculation of the shift due to the Zemach radius $\tilde r_Z$ is more complicated. 
$\tilde r_Z$ represents the hyperfine anomaly, namely $(E_\mathrm{hfs}^\mathrm{exp}-E_\mathrm{hfs}^\mathrm{point})/E_F = -2\,Z\,\alpha\,m\,\tilde r_Z$, see Eq. (\ref{50}).
If we assume that it comes exclusively from the
charge and magnetic moment distribution, it becomes $r_Z$ given by Eq. (\ref{51}),
which can only be derived from the Dirac equation.
Let us thus start derivation from the relativistic hyperfine splitting
\begin{align}
E_\mathrm{hfs} =&\ -e\,\Bigl\langle\psi^\dag\Big|\vec\alpha\cdot\vec A_\mathrm{I}\Big|\psi\Bigr\rangle, \label{44}
\end{align}
where, for a point nucleus
\begin{align}
\vec A_\mathrm{I} =&\ \frac{1}{4\,\pi}\,\vec\mu_\mathrm{I}\times\frac{\vec r}{r^3}. 
\label{45}
\end{align}
In the nonrelativistic limit $E_\mathrm{hfs}$  is given by the Fermi formula
\begin{align}
E_F =&\ -\frac{e}{2\,m}\, \big\langle\phi\big|\big\{\vec\sigma\cdot\vec p\,,\,\vec\sigma\cdot\vec A_\mathrm{I}\big\}\big|\phi\big\rangle
\nonumber \\
=& -\big\langle\phi |\vec\mu_e\cdot\vec B_\mathrm{I} |\phi\big\rangle\nonumber \\
=& -\frac{2}{3}\big\langle\phi |\vec\mu_e\cdot\vec \mu_\mathrm{I}\,\delta^3(r)|\phi\big\rangle. \label{46}
\end{align}
We are now ready to consider the leading finite nuclear size correction $E_Z$ to the hyperfine splitting
\begin{align}
E_Z =&\ 2\,\biggl\langle\! \begin{array}{c}\phi^\dag(0)\\ 0 \end{array}\! \bigg|
(-e)\,\vec\gamma\cdot\vec A_\mathrm{I}\,\frac{1}{\not\!p-m}\,e\,\gamma^0\,A^0
\bigg| \! \begin{array}{c}\phi(0)\\ 0 \end{array}\! \biggr\rangle
\nonumber \\
=&\ 2\,e^2 \int\frac{d^3p}{(2\,\pi)^3}\,\frac{1}{\vec p^{\,2}}\nonumber \\ \times &\ 
\biggl\langle\! \begin{array}{c}\phi^\dag(0)\\ 0 \end{array}\! \bigg|
\vec\gamma\cdot\vec A_\mathrm{I}(-\vec p)\,(\not\!p+m)\,\gamma^0\,A^0(\vec p)
\bigg| \! \begin{array}{c}\phi(0)\\ 0 \end{array}\! \biggr\rangle, \label{47}
\end{align}
where $p^0=m$ and
\begin{align}
A^0(\vec p) =&\ -\frac{Z\,e}{\vec p^{\,2}}\,G_E(\vec p^{\,2}), \label{48}\\
\vec A_\mathrm{I}(\vec p) =&\ -i\, \vec\mu_\mathrm{I}\times\frac{\vec p}{p^2}\,G_M(\vec p^{\,2}), \label{49}
\end{align}
with normalization $G_E(0)=G_M(0) = 1$. $E_Z$ can be simplified to
\begin{align}
E_{\rm Z} =&\ \frac{2\,Z\,\alpha\,m}{\pi^2}\,
\int\frac{d^3p}{p^4}\,\biggl[G_E(p^2)\,G_M(p^2)-1\biggr]\,E_F \nonumber \\
=&\  -2\,Z\,\alpha\, m\,r_{\rm Z}\,E_F
\,,\label{50}
\end{align}
where
\begin{equation}
r_{\rm Z} = \int d^3 r_1 \int d^3 r_2\,\rho_E(r_1)\,\rho_M(r_2)\,|\vec r_1-\vec r_2|, \label{51}
\end{equation}
and where $\rho_E$ and $\rho_M$ are the Fourier transforms of $G_E$ and $G_M$.
If we are about to represent complete hyperfine anomaly, then $r_Z$ becomes $\tilde r_Z$ in Eq. (\ref{50}),
because it may include the nuclear inelastic contribution.

Let us now combine perturbation due to $r_Z$ and the homogenous magnetic field
\begin{align}
\delta E =&\ 2\,\Bigl\langle\!\bar\psi_M \Big|
(-e)\,\vec\gamma\cdot\vec A_\mathrm{I}\,\frac{1}{\not\!p-e\not\!\!A-m}\,e\,\gamma^0\,A^0 \Big| \psi_M \Bigr\rangle, \label{52}
\end{align}
where
 \begin{equation}
  |\psi_M\rangle = \biggl(I-\frac{1}{2\,m}\,\vec\gamma\,(\vec p-e\,\vec A) +\frac{e}{8\,m^2}\,\vec\sigma\vec B\biggr)
  \left|\begin{array}{c} \phi_M(0)\\ 0 \end{array}\right\rangle, \label{53}
  \end{equation}
 and  where $\phi_M$ is an eigenstate of 
  \begin{align}
H_M =&  \frac{p^2}{2\,m} -\frac{Z\,\alpha}{r} -\frac{e}{2\,m}\,\vec\sigma\vec B
  \biggl(1-\frac{p^2}{2\,m^2}+\frac{Z\alpha}{6\,m\,r}\!\biggr). \label{54}
\end{align}
We claim that the $e\,\vec A$ terms in the propagator and in the wave function can be neglected,
because they lead to an additional $p^2$ in the denominator and their contribution thus goes with the nuclear radius to the third power.
Therefore, we have only two corrections due to the last terms in Eqs. (\ref{53}) and (\ref{54}), namely, 
\begin{align}
\delta E =&\ 2\,\langle\phi| H_Z\,\frac{e}{8\,m^2}\,\vec\sigma\vec B\rangle 
\nonumber \\ &\hspace*{-3ex}
- 2\,\Big\langle\phi\Big| H_Z\,\frac{1}{(E-H)'} \frac{e}{2\,m}\,\vec\sigma\vec B
  \biggl(-\frac{p^2}{2\,m^2}+\frac{Z\alpha}{6\,m\,r}\!\biggr)
  \Big|\phi\Big\rangle, \label{55}
\end{align}
where
\begin{align}
H_Z =&\ \frac{2}{3}\vec\mu_e\cdot\vec\mu_\mathrm{I}\,\delta^3(r)\,(2\,Z\,\alpha\,m\,r_Z). \label{56}
\end{align}
Therefore,
\begin{align}
\delta E =&\ \vec\mu_\mathrm{I}\cdot\vec B\,\alpha\,(Z\,\alpha)^3\,m\,r_Z\,\frac{8}{3}\,\biggl(\frac{1}{4} - \frac{X}{m^2\,(Z\,\alpha)^3}\biggr), \label{57}
\end{align}
where
\begin{align}
X=&\ \Big\langle\phi\Big|\pi\,\delta^3(r)\,\frac{1}{(E-H)'} 
  \biggl(-\frac{p^2}{2\,m^2}+\frac{Z\alpha}{6\,m\,r}\!\biggr)
  \Big|\phi\Big\rangle
\nonumber \\ =&\ 
\frac{5}{4}\,(Z\,\alpha)^3\,m^2. \label{58}
\end{align}
Thus with $ \delta E = \vec\mu_\mathrm{I}\cdot\vec B\,\sigma_Z$
\begin{align}
\sigma_Z =&\ - \frac{8}{3}\, \alpha\,(Z\,\alpha)^4\, m\,r_Z, \label{60}
\end{align}
which proves  Eq. (\ref{38}). In addition, Yerokhin \cite{private2} verified this equation by 
numerically calculating the magnetic shielding with Dirac wave functions
for various $Z$, charge, and magnetic radii of the nucleus.
The advantage of Eq. (\ref{38}) over the direct numerical calculation is the presence of $\tilde r_Z$ instead of $r_Z$,
which represents the sum of elastic and inelastic contributions to HFS, and thus can be determined from the HFS anomaly.

\section{Summary}
\newcommand{\mct}[1]{\multicolumn{2}{l}{#1}}
\begin{table}
\caption{Contributions to the shielding constant $10^6\,\sigma$ for $^1$H, $^3$He$^+$, and $^3$He using Ref. \cite{rudzinski09, wehrli:2022}.
New results are $\sigma^{(2,1)}$(He), $\sigma^{(2,2)}$, $\sigma^{(6)}$ and $\sigma_\mathrm{fs}$.
Because  the direct numerical calculation of QED corrections to $\sigma^{(6)}$
is not sufficiently accurate for low $Z$ \cite{yerokhin11,yerokhin12}, we estimate uncertainty from QED corrections at this order by 
assuming that it does not exceed the known relativistic contribution to $\sigma^6$.
$\sigma_\mathrm{fs}$ was calculated using: $r_C(p) = r_M(p) = 0.84$ fm \cite{pohl:10}, $\tilde r_Z(p) = 0.87$ fm \cite{SGK05}, 
$r_C(h) = r_M(h) = 1.97$ fm \cite{schuhmann23}, $\tilde r_Z(h) = 2.60$ fm,
other physical constants are from \cite{tiesinga21}.}
\label{tab:shieldingstot}
\begin{center}
\begin{tabular}{ld{2.12}d{2.12}d{2.12}}
    \hline \hline & \\
 & \centt{$^1$H} &  \centt{$^3$He$^+$} & \centt{$^3$He} \\[1ex]
\hline
$\sigma^{(2,0)}$ &17.750\,451\,5 & 35.500\,903\,0 &  59.936\,771\,0  \\
$\sigma^{(2,1)}$ & -0.017\,603\,7& -0.013\,933\,4 &  -0.023\,020\,1  \\
$\sigma^{(2,2)}$ &  0.000\,014\,1& 0.000\,001\,4& 0.000\,002\,1(7)   \\
$\sigma^{(4,0)}$ &0.002\,546\,9 &  0.020\,375\,1 &   0.052\,663\,1  \\
$\sigma^{(4,1)}$ &  0.000\,000\,0(28) & 0.000\,000\,0(74)& 0.000\,000\,0(192)   \\
$\sigma^{(5,0)}$ &0.000\,018\,4 &  0.000\,082\,0 &   0.000\,096\,3\\
$\sigma^{(6,0)}$ & 0.000\,000\,2(2) & 0.000\,006\,5(65)  & 0.000\,012\,9(129)  \\
$\sigma_\mathrm{fs}$ & -0.000\,000\,1 & -0.000\,006\,7  & -0.000\,013\,5(67)  \\[2ex]
$10^6\,\sigma$& 17.735\,427(3) & 35.507\,427(10) &   59.966\,512(24)\\
Previous                 & 17.735\,436(3) & 35.507\,434(9) &   59.967\,029(23)\\
\hline\hline& 
\end{tabular}
\end{center}
\end{table}

The total magnetic shielding for hydrogen-like ions including contributions up to order $\alpha^6$ is (cf. Eq. (25) of Ref. \cite{wehrli:2021})
\begin{align}
\sigma =&\ \frac{Z\,\alpha^2}{3}+\frac{97}{108}\,Z^3\,\alpha^4 + \frac{289}{216}\,Z^5\,\alpha^6
+ \frac{8\,\alpha^2}{9\,\pi}\,(Z\,\alpha)^3
\nonumber\\ & \times
\biggl[\ln(Z\,\alpha)^{-2} +2\ln k_0-3\ln k_3- \frac{221}{64}+\frac{3}{5}\biggr] \nonumber \\
&+ \frac{Z\,\alpha^2}{3}\,\biggl[\biggl( \frac{1}{g_N}  - 2\biggr)\frac{m}{m_N}
+\bigg(\frac{4\,Z-3}{g_N} - \frac{Z}{2} + 3 \bigg)\frac{m^2}{m_N^2}\biggr] \nonumber\\
 &\ -\frac{Z\,\alpha^2}{3}\,\bigl[
  2\,(Z\,\alpha)^2\,m^2\,(r_C^2+r_M^2) +8\,(Z\,\alpha)^3\,m\,\tilde r_Z \bigr], \label{61}
\end{align}
where \cite{pachucki05}
\begin{align}
\ln k_0 =&\ 2.984\,128\,556,\\
\ln k_3 =&\ 3.272\,806\,545\,.
\end{align}
Numerical results for all these known contributions to the magnetic shielding in H, He$^+$, and He are presented in Table I. 
The updated values are $\sigma^{(2,1)}$ for He, where we corrected the sign error in the last term in Eq. (\ref{32}).
This leading recoil correction to the magnetic shielding  is about $0.02\cdot 10^{-6}$, which is the relative $2\cdot 10^{-8}$ 
correction in the determination of nuclear magnetic moments. The higher-order recoil correction,  the last term in Eq. (\ref{37}),
which is also corrected in this work, is much smaller and thus is negligible at present accuracy of measurements.
The same holds for nuclear finite size effects, described by Eq.  (\ref{38}); they are negligible for light elements
and can safely be neglected.
However, the nuclear finite size effects can be significant for heavy elements, where  they strongly affect binding energies and hyperfine splitting.  

Finally, our recommended values for the nuclear magnetic shieldings are in the  penultimate row, 
and they are compared to previous recommendations from Ref. \cite{wehrli:2022} in the last row. 
The largest change of $0.5\cdot 10^{-9}$ is for the He atom; changes to H and He$^+$ ion are negligible.

We can now use these new shieldings to recalculate the helion magnetic moment from He$^+$ measurement
\begin{align}
\mu(^3\mathrm{He}^+) =&\  -4.255\,099\,606\,9(30)(17)\times\frac{\mu_N}{2}, \label{01}
\end{align}  
namely, it is
\begin{align}
\mu(^3\mathrm{He}^{++}) =&\ \frac{\mu(^3\mathrm{He}^+)} {1-\sigma(^3\mathrm{He}^+)} 
\nonumber \\ =&\  -2.127\,625\,350\,0(17)\,\mu_N\,, \label{64}
\end{align}
which  differs slightly from that in  Ref. \cite{schneider2022} , while  our recommended value for the atomic $^3$He magnetic moment is 
\begin{align}
\mu(^3\mathrm{He}) =&\   \mu(^3\mathrm{He}^+) \frac{1-\sigma(^3\mathrm{He})} {1-\sigma(^3\mathrm{He}^+)} 
\nonumber \\ =&\   -2.127\,497\,763\,7(17)\,\mu_N\,, \label{65}
\end{align}
which can serve as a reference in gaseous NMR measurements \cite{Gentile:2017} because it is the most accurately known
atomic magnetic moment.

\begin{acknowledgments}
We wish to thank Vladimir Yerokhin for numerical calculations of the finite nuclear size effects
in hydrogen-like ions, and Jan K\l os for participation at the beginning of the project. This work was
supported by the National Science Center (Poland) Grant No. 2017/27/B/ST2/02459.
\end{acknowledgments}


\begin{thebibliography}{18}%
\makeatletter
\providecommand \@ifxundefined [1]{%
 \@ifx{#1\undefined}
}%
\providecommand \@ifnum [1]{%
 \ifnum #1\expandafter \@firstoftwo
 \else \expandafter \@secondoftwo
 \fi
}%
\providecommand \@ifx [1]{%
 \ifx #1\expandafter \@firstoftwo
 \else \expandafter \@secondoftwo
 \fi
}%
\providecommand \natexlab [1]{#1}%
\providecommand \enquote  [1]{``#1''}%
\providecommand \bibnamefont  [1]{#1}%
\providecommand \bibfnamefont [1]{#1}%
\providecommand \citenamefont [1]{#1}%
\providecommand \href@noop [0]{\@secondoftwo}%
\providecommand \href [0]{\begingroup \@sanitize@url \@href}%
\providecommand \@href[1]{\@@startlink{#1}\@@href}%
\providecommand \@@href[1]{\endgroup#1\@@endlink}%
\providecommand \@sanitize@url [0]{\catcode `\\12\catcode `\$12\catcode
  `\&12\catcode `\#12\catcode `\^12\catcode `\_12\catcode `\%12\relax}%
\providecommand \@@startlink[1]{}%
\providecommand \@@endlink[0]{}%
\providecommand \url  [0]{\begingroup\@sanitize@url \@url }%
\providecommand \@url [1]{\endgroup\@href {#1}{\urlprefix }}%
\providecommand \urlprefix  [0]{URL }%
\providecommand \Eprint [0]{\href }%
\providecommand \doibase [0]{https://doi.org/}%
\providecommand \selectlanguage [0]{\@gobble}%
\providecommand \bibinfo  [0]{\@secondoftwo}%
\providecommand \bibfield  [0]{\@secondoftwo}%
\providecommand \translation [1]{[#1]}%
\providecommand \BibitemOpen [0]{}%
\providecommand \bibitemStop [0]{}%
\providecommand \bibitemNoStop [0]{.\EOS\space}%
\providecommand \EOS [0]{\spacefactor3000\relax}%
\providecommand \BibitemShut  [1]{\csname bibitem#1\endcsname}%
\let\auto@bib@innerbib\@empty
\bibitem [{\citenamefont {Schneider}\ \emph {et~al.}(2022)\citenamefont
  {Schneider}, \citenamefont {Sikora}, \citenamefont {Dickopf}, \citenamefont
  {M{\"u}ller}, \citenamefont {Oreshkina}, \citenamefont {Rischka},
  \citenamefont {Valuev}, \citenamefont {Ulmer}, \citenamefont {Walz},
  \citenamefont {Harman}, \citenamefont {Keitel}, \citenamefont {Mooser},\ and\
  \citenamefont {Blaum}}]{schneider2022}%
  \BibitemOpen
  \bibfield  {author} {\bibinfo {author} {\bibfnamefont {A.}~\bibnamefont
  {Schneider}}, \bibinfo {author} {\bibfnamefont {B.}~\bibnamefont {Sikora}},
  \bibinfo {author} {\bibfnamefont {S.}~\bibnamefont {Dickopf}}, \bibinfo
  {author} {\bibfnamefont {M.}~\bibnamefont {M{\"u}ller}}, \bibinfo {author}
  {\bibfnamefont {N.~S.}\ \bibnamefont {Oreshkina}}, \bibinfo {author}
  {\bibfnamefont {A.}~\bibnamefont {Rischka}}, \bibinfo {author} {\bibfnamefont
  {I.~A.}\ \bibnamefont {Valuev}}, \bibinfo {author} {\bibfnamefont
  {S.}~\bibnamefont {Ulmer}}, \bibinfo {author} {\bibfnamefont
  {J.}~\bibnamefont {Walz}}, \bibinfo {author} {\bibfnamefont {Z.}~\bibnamefont
  {Harman}}, \bibinfo {author} {\bibfnamefont {C.~H.}\ \bibnamefont {Keitel}},
  \bibinfo {author} {\bibfnamefont {A.}~\bibnamefont {Mooser}},\ and\ \bibinfo
  {author} {\bibfnamefont {K.}~\bibnamefont {Blaum}},\ }\href
  {https://doi.org/10.1038/s41586-022-04761-7} {\bibfield  {journal} {\bibinfo
  {journal} {Nature}\ }\textbf {\bibinfo {volume} {606}},\ \bibinfo {pages}
  {878} (\bibinfo {year} {2022})}\BibitemShut {NoStop}%
\bibitem [{\citenamefont {Wehrli}\ \emph {et~al.}(2021)\citenamefont {Wehrli},
  \citenamefont {Spyszkiewicz-Kaczmarek}, \citenamefont {Puchalski},\ and\
  \citenamefont {Pachucki}}]{wehrli:2021}%
  \BibitemOpen
  \bibfield  {author} {\bibinfo {author} {\bibfnamefont {D.}~\bibnamefont
  {Wehrli}}, \bibinfo {author} {\bibfnamefont {A.}~\bibnamefont
  {Spyszkiewicz-Kaczmarek}}, \bibinfo {author} {\bibfnamefont {M.}~\bibnamefont
  {Puchalski}},\ and\ \bibinfo {author} {\bibfnamefont {K.}~\bibnamefont
  {Pachucki}},\ }\href {https://doi.org/10.1103/PhysRevLett.127.263001}
  {\bibfield  {journal} {\bibinfo  {journal} {Phys. Rev. Lett.}\ }\textbf
  {\bibinfo {volume} {127}},\ \bibinfo {pages} {263001} (\bibinfo {year}
  {2021})}\BibitemShut {NoStop}%
\bibitem [{\citenamefont {Wehrli}\ \emph {et~al.}(2022)\citenamefont {Wehrli},
  \citenamefont {Puchalski},\ and\ \citenamefont {Pachucki}}]{wehrli:2022}%
  \BibitemOpen
  \bibfield  {author} {\bibinfo {author} {\bibfnamefont {D.}~\bibnamefont
  {Wehrli}}, \bibinfo {author} {\bibfnamefont {M.}~\bibnamefont {Puchalski}},\
  and\ \bibinfo {author} {\bibfnamefont {K.}~\bibnamefont {Pachucki}},\ }\href
  {https://doi.org/10.1103/PhysRevA.105.032808} {\bibfield  {journal} {\bibinfo
   {journal} {Phys. Rev. A}\ }\textbf {\bibinfo {volume} {105}},\ \bibinfo
  {pages} {032808} (\bibinfo {year} {2022})}\BibitemShut {NoStop}%
\bibitem [{\citenamefont {Mooser}(2021)}]{private3}%
  \BibitemOpen
  \bibfield  {author} {\bibinfo {author} {\bibfnamefont {A.}~\bibnamefont
  {Mooser}},\ }\href@noop {} {}\bibinfo {howpublished} {private communication}
  (\bibinfo {year} {2021})\BibitemShut {NoStop}%
\bibitem [{\citenamefont {Karshenboim}(2005)}]{SGK05}%
  \BibitemOpen
  \bibfield  {author} {\bibinfo {author} {\bibfnamefont {S.~G.}\ \bibnamefont
  {Karshenboim}},\ }\href
  {https://doi.org/https://doi.org/10.1016/j.physrep.2005.08.008} {\bibfield
  {journal} {\bibinfo  {journal} {Physics Reports}\ }\textbf {\bibinfo {volume}
  {422}},\ \bibinfo {pages} {1} (\bibinfo {year} {2005})}\BibitemShut {NoStop}%
\bibitem [{\citenamefont {Jaszu{\'n}ski}\ \emph {et~al.}(2012)\citenamefont
  {Jaszu{\'n}ski}, \citenamefont {{Antu\v sek}}, \citenamefont {Garbacz},
  \citenamefont {Jackowski}, \citenamefont {Makulski},\ and\ \citenamefont
  {Wilczek}}]{jaszunski2012}%
  \BibitemOpen
  \bibfield  {author} {\bibinfo {author} {\bibfnamefont {M.}~\bibnamefont
  {Jaszu{\'n}ski}}, \bibinfo {author} {\bibfnamefont {A.}~\bibnamefont {{Antu\v
  sek}}}, \bibinfo {author} {\bibfnamefont {P.}~\bibnamefont {Garbacz}},
  \bibinfo {author} {\bibfnamefont {K.}~\bibnamefont {Jackowski}}, \bibinfo
  {author} {\bibfnamefont {W.}~\bibnamefont {Makulski}},\ and\ \bibinfo
  {author} {\bibfnamefont {M.}~\bibnamefont {Wilczek}},\ }\href
  {https://doi.org/https://doi.org/10.1016/j.pnmrs.2012.03.002} {\bibfield
  {journal} {\bibinfo  {journal} {Prog. Nucl. Magn. Reson. Spectrosc.}\
  }\textbf {\bibinfo {volume} {67}},\ \bibinfo {pages} {49} (\bibinfo {year}
  {2012})}\BibitemShut {NoStop}%
\bibitem [{\citenamefont {Pachucki}(2008)}]{pachucki08}%
  \BibitemOpen
  \bibfield  {author} {\bibinfo {author} {\bibfnamefont {K.}~\bibnamefont
  {Pachucki}},\ }\href {https://doi.org/10.1103/PhysRevA.78.012504} {\bibfield
  {journal} {\bibinfo  {journal} {Phys. Rev. A}\ }\textbf {\bibinfo {volume}
  {78}},\ \bibinfo {pages} {012504} (\bibinfo {year} {2008})}\BibitemShut
  {NoStop}%
\bibitem [{\citenamefont {Rudzi{\'n}ski}\ \emph {et~al.}(2009)\citenamefont
  {Rudzi{\'n}ski}, \citenamefont {Puchalski},\ and\ \citenamefont
  {Pachucki}}]{rudzinski09}%
  \BibitemOpen
  \bibfield  {author} {\bibinfo {author} {\bibfnamefont {A.}~\bibnamefont
  {Rudzi{\'n}ski}}, \bibinfo {author} {\bibfnamefont {M.}~\bibnamefont
  {Puchalski}},\ and\ \bibinfo {author} {\bibfnamefont {K.}~\bibnamefont
  {Pachucki}},\ }\href {https://doi.org/10.1063/1.3159674} {\bibfield
  {journal} {\bibinfo  {journal} {J. Chem. Phys.}\ }\textbf {\bibinfo {volume}
  {130}},\ \bibinfo {pages} {244102} (\bibinfo {year} {2009})}\BibitemShut
  {NoStop}%
\bibitem [{\citenamefont {Yerokhin}\ \emph {et~al.}(2011)\citenamefont
  {Yerokhin}, \citenamefont {Pachucki}, \citenamefont {Harman},\ and\
  \citenamefont {Keitel}}]{yerokhin11}%
  \BibitemOpen
  \bibfield  {author} {\bibinfo {author} {\bibfnamefont {V.~A.}\ \bibnamefont
  {Yerokhin}}, \bibinfo {author} {\bibfnamefont {K.}~\bibnamefont {Pachucki}},
  \bibinfo {author} {\bibfnamefont {Z.}~\bibnamefont {Harman}},\ and\ \bibinfo
  {author} {\bibfnamefont {C.~H.}\ \bibnamefont {Keitel}},\ }\href
  {https://doi.org/10.1103/PhysRevLett.107.043004} {\bibfield  {journal}
  {\bibinfo  {journal} {Phys. Rev. Lett.}\ }\textbf {\bibinfo {volume} {107}},\
  \bibinfo {pages} {043004} (\bibinfo {year} {2011})}\BibitemShut {NoStop}%
\bibitem [{\citenamefont {Yerokhin}\ \emph {et~al.}(2012)\citenamefont
  {Yerokhin}, \citenamefont {Pachucki}, \citenamefont {Harman},\ and\
  \citenamefont {Keitel}}]{yerokhin12}%
  \BibitemOpen
  \bibfield  {author} {\bibinfo {author} {\bibfnamefont {V.~A.}\ \bibnamefont
  {Yerokhin}}, \bibinfo {author} {\bibfnamefont {K.}~\bibnamefont {Pachucki}},
  \bibinfo {author} {\bibfnamefont {Z.}~\bibnamefont {Harman}},\ and\ \bibinfo
  {author} {\bibfnamefont {C.~H.}\ \bibnamefont {Keitel}},\ }\href
  {https://doi.org/10.1103/PhysRevA.85.022512} {\bibfield  {journal} {\bibinfo
  {journal} {Phys. Rev. A}\ }\textbf {\bibinfo {volume} {85}},\ \bibinfo
  {pages} {022512} (\bibinfo {year} {2012})}\BibitemShut {NoStop}%
\bibitem [{\citenamefont {Hegstrom}(1973)}]{hegstrom73}%
  \BibitemOpen
  \bibfield  {author} {\bibinfo {author} {\bibfnamefont {R.~A.}\ \bibnamefont
  {Hegstrom}},\ }\href {https://doi.org/10.1103/PhysRevA.7.451} {\bibfield
  {journal} {\bibinfo  {journal} {Phys. Rev. A}\ }\textbf {\bibinfo {volume}
  {7}},\ \bibinfo {pages} {451} (\bibinfo {year} {1973})}\BibitemShut {NoStop}%
\bibitem [{\citenamefont {Grotch}\ and\ \citenamefont
  {Hegstrom}(1971)}]{grotch71}%
  \BibitemOpen
  \bibfield  {author} {\bibinfo {author} {\bibfnamefont {H.}~\bibnamefont
  {Grotch}}\ and\ \bibinfo {author} {\bibfnamefont {R.~A.}\ \bibnamefont
  {Hegstrom}},\ }\bibfield  {journal} {\bibinfo  {journal} {Phys. Rev. A}\
  }\href {https://doi.org/10.1103/PhysRevA.4.59} {10.1103/PhysRevA.4.59}
  (\bibinfo {year} {1971})\BibitemShut {NoStop}%
\bibitem [{\citenamefont {Yerokhin}(2023)}]{private2}%
  \BibitemOpen
  \bibfield  {author} {\bibinfo {author} {\bibfnamefont {V.}~\bibnamefont
  {Yerokhin}},\ }\href@noop {} {}\bibinfo {howpublished} {private
  communication} (\bibinfo {year} {2023})\BibitemShut {NoStop}%
\bibitem [{\citenamefont {Pohl}\ \emph {et~al.}(2010)\citenamefont {Pohl},
  \citenamefont {Antognini}, \citenamefont {Nez}, \citenamefont {Amaro},
  \citenamefont {Biraben}, \citenamefont {Cardoso}, \citenamefont {Covita},
  \citenamefont {Dax}, \citenamefont {Dhawan}, \citenamefont {Fernandes},
  \citenamefont {Giesen}, \citenamefont {Graf}, \citenamefont {H\"ansch},
  \citenamefont {Indelicato}, \citenamefont {Julien}, \citenamefont {Kao},
  \citenamefont {Knowles}, \citenamefont {Le~Bigot}, \citenamefont {Liu},
  \citenamefont {Lopes {\em et al.}}}]{pohl:10}%
  \BibitemOpen
  \bibfield  {author} {\bibinfo {author} {\bibfnamefont {R.}~\bibnamefont
  {Pohl}}, \bibinfo {author} {\bibfnamefont {A.}~\bibnamefont {Antognini}},
  \bibinfo {author} {\bibfnamefont {F.}~\bibnamefont {Nez}}, \bibinfo {author}
  {\bibfnamefont {F.~D.}\ \bibnamefont {Amaro}}, \bibinfo {author}
  {\bibfnamefont {F.}~\bibnamefont {Biraben}}, \bibinfo {author} {\bibfnamefont
  {J.~a. M.~R.}\ \bibnamefont {Cardoso}}, \bibinfo {author} {\bibfnamefont
  {D.~S.}\ \bibnamefont {Covita}}, \bibinfo {author} {\bibfnamefont
  {A.}~\bibnamefont {Dax}}, \bibinfo {author} {\bibfnamefont {S.}~\bibnamefont
  {Dhawan}}, \bibinfo {author} {\bibfnamefont {L.~M.~P.}\ \bibnamefont
  {Fernandes}}, \bibinfo {author} {\bibfnamefont {A.}~\bibnamefont {Giesen}},
  \bibinfo {author} {\bibfnamefont {T.}~\bibnamefont {Graf}}, \bibinfo {author}
  {\bibfnamefont {T.~W.}\ \bibnamefont {H\"ansch}}, \bibinfo {author}
  {\bibfnamefont {P.}~\bibnamefont {Indelicato}}, \bibinfo {author}
  {\bibfnamefont {L.}~\bibnamefont {Julien}}, \bibinfo {author} {\bibfnamefont
  {C.-Y.}\ \bibnamefont {Kao}}, \bibinfo {author} {\bibfnamefont
  {P.}~\bibnamefont {Knowles}}, \bibinfo {author} {\bibfnamefont {E.-O.}\
  \bibnamefont {Le~Bigot}}, \bibinfo {author} {\bibfnamefont {Y.-W.}\
  \bibnamefont {Liu}}, \bibinfo {author} {\bibfnamefont {J.~A.~M.}\
  \bibnamefont {Lopes {\em et al.}, }}}\href@noop {} {\bibfield  {journal}
  {\bibinfo  {journal} {Nature (London)}\ }\textbf {\bibinfo {volume} {466}},\
  \bibinfo {pages} {213 } (\bibinfo {year} {2010})}\BibitemShut {NoStop}%
\bibitem [{\citenamefont {Collaboration}\ \emph {et~al.}(2023)\citenamefont
  {Collaboration}, \citenamefont {Schuhmann}, \citenamefont {Fernandes},
  \citenamefont {Nez}, \citenamefont {Ahmed}, \citenamefont {Amaro},
  \citenamefont {Amaro}, \citenamefont {Biraben}, \citenamefont {Chen},
  \citenamefont {Covita}, \citenamefont {Dax}, \citenamefont {Diepold},
  \citenamefont {Franke}, \citenamefont {Galtier}, \citenamefont {Gouvea},
  \citenamefont {Götzfried}, \citenamefont {Graf}, \citenamefont {Hänsch},
  \citenamefont {Hildebrandt}, \citenamefont {Indelicato}, \citenamefont
  {Julien et al}}]{schuhmann23}%
  \BibitemOpen
  \bibfield  {author} {\bibinfo {author} {\bibfnamefont {K.}~\bibnamefont
  {Schuhmann}}, \bibinfo {author} {\bibfnamefont {L.~M.~P.}\ \bibnamefont
  {Fernandes}}, \bibinfo {author} {\bibfnamefont {F.}~\bibnamefont {Nez}},
  \bibinfo {author} {\bibfnamefont {M.~A.}\ \bibnamefont {Ahmed}}, \bibinfo
  {author} {\bibfnamefont {F.~D.}\ \bibnamefont {Amaro}}, \bibinfo {author}
  {\bibfnamefont {P.}~\bibnamefont {Amaro}}, \bibinfo {author} {\bibfnamefont
  {F.}~\bibnamefont {Biraben}}, \bibinfo {author} {\bibfnamefont {T.-L.}\
  \bibnamefont {Chen}}, \bibinfo {author} {\bibfnamefont {D.~S.}\ \bibnamefont
  {Covita}}, \bibinfo {author} {\bibfnamefont {A.~J.}\ \bibnamefont {Dax}},
  \bibinfo {author} {\bibfnamefont {M.}~\bibnamefont {Diepold}}, \bibinfo
  {author} {\bibfnamefont {B.}~\bibnamefont {Franke}}, \bibinfo {author}
  {\bibfnamefont {S.}~\bibnamefont {Galtier}}, \bibinfo {author} {\bibfnamefont
  {A.~L.}\ \bibnamefont {Gouvea}}, \bibinfo {author} {\bibfnamefont
  {J.}~\bibnamefont {Götzfried}}, \bibinfo {author} {\bibfnamefont
  {T.}~\bibnamefont {Graf}}, \bibinfo {author} {\bibfnamefont {T.~W.}\
  \bibnamefont {Hänsch}}, \bibinfo {author} {\bibfnamefont {M.}~\bibnamefont
  {Hildebrandt}}, \bibinfo {author} {\bibfnamefont {P.}~\bibnamefont
  {Indelicato}}, \bibinfo {author} {\bibfnamefont {L.}~\bibnamefont {Julien {\em et al.}}},
  }\href@noop {}  (\bibinfo {year} {2023}),\
  \Eprint {https://arxiv.org/abs/2305.11679} {arXiv:2305.11679
  [physics.atom-ph]} \BibitemShut {NoStop}%
\bibitem [{\citenamefont {Tiesinga}\ \emph {et~al.}(2021)\citenamefont
  {Tiesinga}, \citenamefont {Mohr}, \citenamefont {Newell},\ and\ \citenamefont
  {Taylor}}]{tiesinga21}%
  \BibitemOpen
  \bibfield  {author} {\bibinfo {author} {\bibfnamefont {E.}~\bibnamefont
  {Tiesinga}}, \bibinfo {author} {\bibfnamefont {P.~J.}\ \bibnamefont {Mohr}},
  \bibinfo {author} {\bibfnamefont {D.~B.}\ \bibnamefont {Newell}},\ and\
  \bibinfo {author} {\bibfnamefont {B.~N.}\ \bibnamefont {Taylor}},\ }\href
  {https://doi.org/10.1103/RevModPhys.93.025010} {\bibfield  {journal}
  {\bibinfo  {journal} {Rev. Mod. Phys.}\ }\textbf {\bibinfo {volume} {93}},\
  \bibinfo {pages} {025010} (\bibinfo {year} {2021})}\BibitemShut {NoStop}%
\bibitem [{\citenamefont {Pachucki}\ \emph {et~al.}(2005)\citenamefont
  {Pachucki}, \citenamefont {Czarnecki}, \citenamefont {Jentschura},\ and\
  \citenamefont {Yerokhin}}]{pachucki05}%
  \BibitemOpen
  \bibfield  {author} {\bibinfo {author} {\bibfnamefont {K.}~\bibnamefont
  {Pachucki}}, \bibinfo {author} {\bibfnamefont {A.}~\bibnamefont {Czarnecki}},
  \bibinfo {author} {\bibfnamefont {U.~D.}\ \bibnamefont {Jentschura}},\ and\
  \bibinfo {author} {\bibfnamefont {V.~A.}\ \bibnamefont {Yerokhin}},\ }\href
  {https://doi.org/10.1103/PhysRevA.72.022108} {\bibfield  {journal} {\bibinfo
  {journal} {Phys. Rev. A}\ }\textbf {\bibinfo {volume} {72}},\ \bibinfo
  {pages} {022108} (\bibinfo {year} {2005})}\BibitemShut {NoStop}%
\bibitem [{\citenamefont {Gentile}\ \emph {et~al.}(2017)\citenamefont
  {Gentile}, \citenamefont {Nacher}, \citenamefont {Saam},\ and\ \citenamefont
  {Walker}}]{Gentile:2017}%
  \BibitemOpen
  \bibfield  {author} {\bibinfo {author} {\bibfnamefont {T.~R.}\ \bibnamefont
  {Gentile}}, \bibinfo {author} {\bibfnamefont {P.~J.}\ \bibnamefont {Nacher}},
  \bibinfo {author} {\bibfnamefont {B.}~\bibnamefont {Saam}},\ and\ \bibinfo
  {author} {\bibfnamefont {T.~G.}\ \bibnamefont {Walker}},\ }\href
  {https://doi.org/10.1103/RevModPhys.89.045004} {\bibfield  {journal}
  {\bibinfo  {journal} {Rev. Mod. Phys.}\ }\textbf {\bibinfo {volume} {89}},\
  \bibinfo {pages} {045004} (\bibinfo {year} {2017})}\BibitemShut {NoStop}%
\end{thebibliography}

%

\end{document}